# Electron-beam induced emergence of mesoscopic ordering in layered MnPS$_3$

Kevin M. Roccapriore,[1*] Nan Huang,[2] Mark P. Oxley,[1] Vinit Sharma,[4,5] Timothy Taylor,[6] Swagata Acharya,[7] Dimitar Pashov,[8] Mikhail I. Katsnelson,[7] David Mandrus,[2,3] Janice L. Musfeldt,[6,9] and Sergei V. Kalinin[1*]

[1] The Center for Nanophase Materials Sciences, Oak Ridge National Laboratory, Oak Ridge, TN 37831, USA

[2] Department of Materials Science and Engineering, University of Tennessee, Knoxville TN, 37916, USA

[3] Materials Science and Technology Division, Oak Ridge National Laboratory, Oak Ridge, TN 37831, USA

[4] Computational Sciences Division, Oak Ridge National Laboratory, Oak Ridge, TN 37831, USA

[5] National Institute for Computational Sciences, University of Tennessee, Knoxville TN, 37831, USA

[6] Department of Chemistry, University of Tennessee, Knoxville, Tennessee 37996, USA

[7] Institute for Molecules and Materials, Radboud University, NL-6525 AJ Nijmegen, The Netherlands

[8] King's College London, Theory and Simulation of Condensed Matter, The Strand, WC2R 2LS London, UK

[9] Department of Physics and Astronomy, University of Tennessee, Knoxville, Tennessee 37996, USA

ABSTRACT

Ordered mesoscale structures in 2D materials induced by small misorientations have allowed for a wide variety of electronic, ferroelectric, and quantum phenomena to be explored. Until now, the only mechanism to induce this periodic ordering was via mechanical rotations between the layers, with the periodicity of the resulting moiré pattern being directly related to twist angle. Here we report a fundamentally distinct mechanism for emergence of mesoscopic periodic patterns in multilayer sulfur-containing metal phosphorous trichalcogenide, MnPS$_3$, induced by the electron beam. The formation under the beam of periodic hexagonal patterns with several characteristic length scales, nucleation and transitions between the phases, and local dynamics are demonstrated.

---

[*] roccapriorkm@ornl.gov
[*] sergei2@utk.edu

The associated mechanisms are attributed to the relative contraction of the layers caused by beam-induced sulphur vacancy formation with subsequent ordering and lattice parameter change. As a result, the plasmonic response of the system is locally altered, suggesting an element of control over plasmon resonances by electron beam patterning. We pose that harnessing this phenomenon provides both insight into fundamental physics of quantum materials and enables device applications by enabling controlled periodic potentials on the atomic scale.

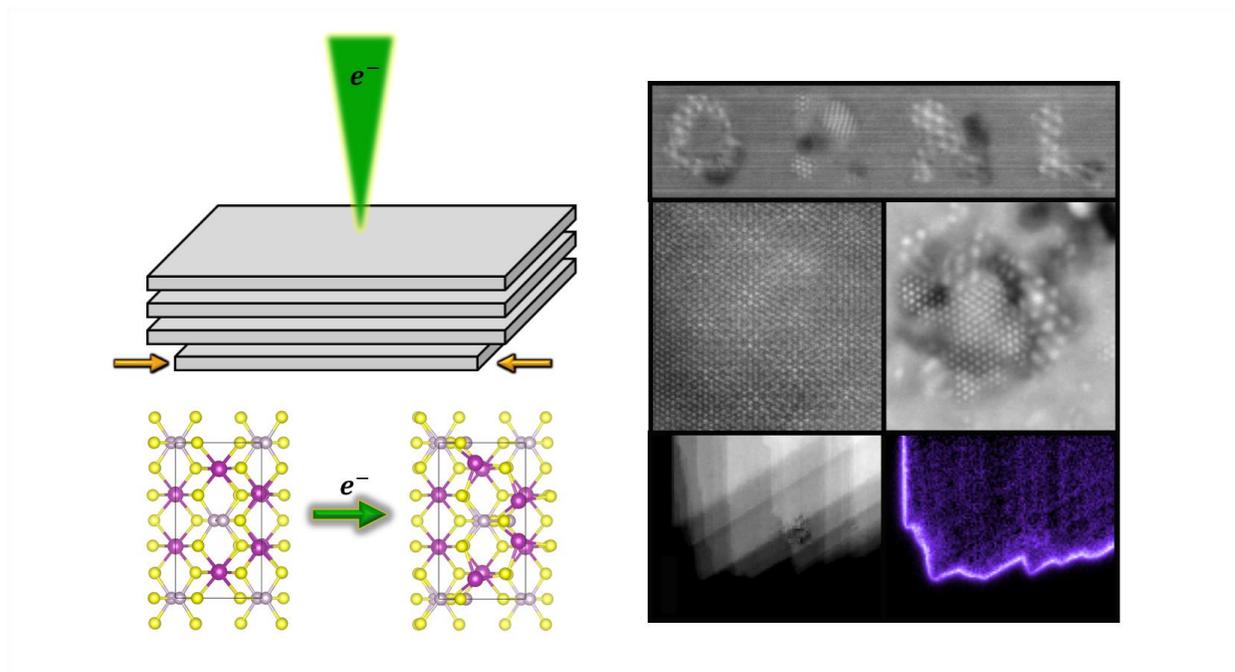

KEYWORDS

moiré superlattice; two-dimensional semiconductors; metal phosphorus trichalcogenides; electron irradiation; edge plasmon; atomic defects; scanning transmission electron microscopy

Over the last 15 years, two-dimensional materials including graphene, boron nitride, and transition metal dichalcogenides have propelled to the forefront of scientific research due to the combination of structural, electronic, and quantum properties they offer. Shrinking the dimensionality of a bulk crystal down to two dimensions invariably leads to unconventional physics by imposing confinement effects, creating anomalous structural configurations, and enabling the emergence of unusual electronic structure. Owing to the incredible ease of preparing a few or even a single layer of atoms, many two-dimensional materials offer a superb platform to rapidly conduct an enormous variety of experiments.

While the initial progress was based preponderantly on device-level measurements, advances in scanning transmission electron microscopy (STEM) have ushered an era in understanding local structural, electronic, and even vibrational properties of materials at the atomic scale. The structure, property, and dynamic of individual point defects, defect complexes, and grain boundaries have been extensively explored via aberration-corrected structural imaging. The development of monochromated[1] high resolution electron energy loss spectroscopy has enabled direct mapping of plasmons and excitons in these materials.[2] The 4D-STEM techniques[3] have enabled high precision structural mapping,[4] allowing exploration of strain and symmetry breaking phenomena,[5] as well probing the detailed 3D structure of defects. Finally, the potential of the electron beam to induce controllable chemical changes has been harnessed to create controlled edges,[6] atomically-defined nanopores and device structures,[7] and even controllably move individual dopant atoms and assemble homo- and heteroatomic artificial molecules.[8,9]

Over the last three years, the field has developed the impetus with the discovery of a wealth of phenomena in twisted van der Waals (vdW) heterostructures.[10] Here, the discovery of superconductivity in twisted bilayer graphene[11,12] and more recently twisted trilayer graphene[13,14] has sparked a wave of interest in these materials. Immediately afterwards was the realization that ferroelectricity is also possible within the twisted material paradigm in graphene and boron nitride.[15] Similarly, STEM has offered a set of insights into atomistic mechanisms, demonstrating the presence of topological defects that may give rise to effects such as second harmonic generation, strain solitons, and adsorption enhancement.[16–18] Jointly, these developments have seeded the field of twistronics and have significantly exploited the functionalities and behaviors manifest in 2D materials, further enabling a broad set of device-relevant applications.

However, this broad variety of emergent functionalities is ultimately derived based on the mechanical rotation of two or more layers with respect to each other. This constraint both necessitates complex fabrication routines and renders the average rotation angle a conserved parameter. While topological defects and ferroelectric domains can be manipulated, the background rotation angle can traditionally be controlled only mechanically by scanning probe or stamping[19–21], and only recently was it discovered to be possible using focused lasers.[22] Additional methods of manipulation are needed to fully explore the set of properties arising within moiré materials. Hence, of significant interest is the development of mesoscopic ordered structures allowing for control of local potential.

Here, we report the emergence of periodically ordered structures and their manipulation using the electron beam in $MnPS_3$, a lesser studied complex vdW material. This material belongs to the family of metal phosphorous trichalcogenides,[23,24] and has been considered as a potential candidate material with non-trivial 2D magnetic ordering. Similar to many 2D materials, $MnPS_3$ can be easily exfoliated to yield flakes of single or several layers. These to our knowledge have only recently begun to be explored for possible functionalities of interest, however comparison with other metal phosphorous trichalcogenides suggest potential emergence of ferro- and antiferroelectric behavior, ionic conduction, and other functionalities. Similarly, this material can serve as a host for cationic substitution on the metal site, and anion substitution on the chalcogen site. Curiously, the electronic properties of bulk and single and few layers were investigated by EELS but with parallel illumination in the TEM,[25,26] but no lattice compression or otherwise peculiar structural findings were noted. The antiferromagnetic transition has been studied by both Raman spectroscopy[27,28] and tunneling magnetoresistance measurements[29,30] as long as fifty years ago. Aside from magnetism approaching the two-dimensional limit, $MnPS_3$ has also been considered a candidate for transistors, UV photodetectors,[31] $NO_2$ gas sensing,[32] and battery electrodes with Li intercalation.[33] More recently, the vibrational properties of $MnPS_3$ were studied with near-field infrared spectroscopy where layer-dependent symmetry crossovers were realized.[34,35] $MnPS_3$ also was found to display piezochromism.[36] Lastly, vacancy formation – which is of crucial significance to the present work – and its effect on the electrical, magnetic, and optical properties of $MnPS_3$ has also been previously investigated with first principles studies.[37]

It is important to mention that many of the compounds in the metal phosphorous trichalcogenide family are notorious for being beam sensitive to STEM imaging, presumably due to sulfur being easily removed at a variety of accelerating voltages, primarily a result of knock-on damage or radiolysis.[38,39] In fact, the recent availability of low acceleration voltage STEM enabled by aberration correction was in part the motivation behind this study.

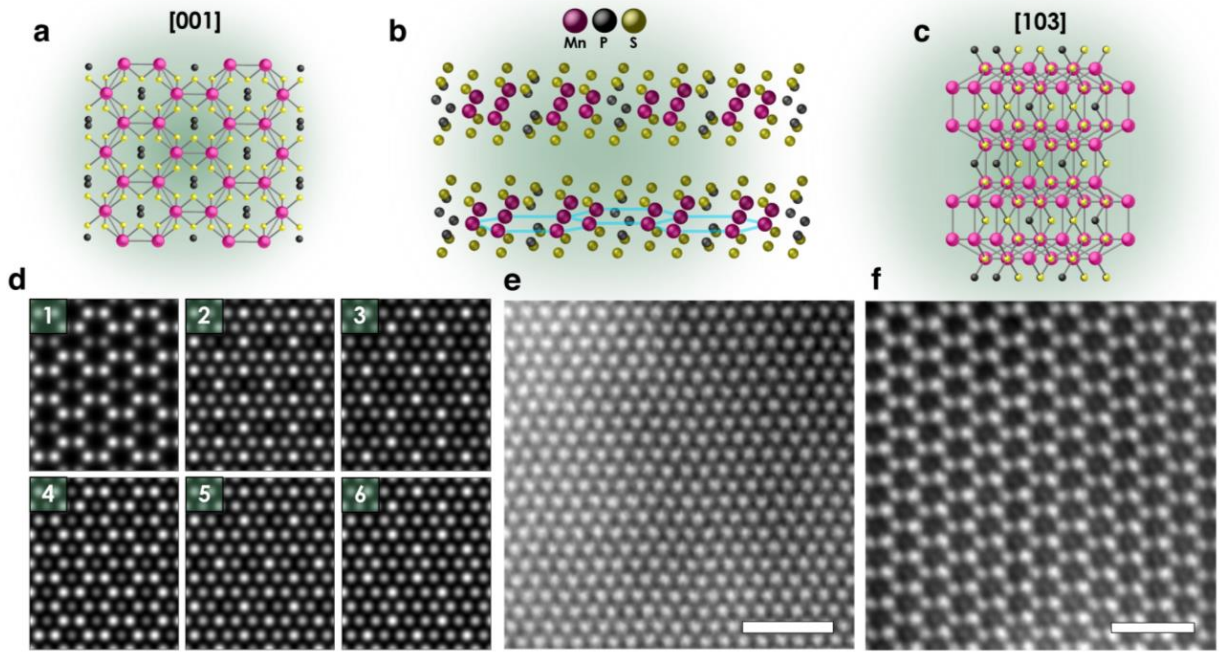

**Figure 1**. MnPS$_3$ structure and layer dependence. Two selected orientations shown in **a** and **c** with 3D view in **b**, where Mn hexagons are highlighted for the bottom layer. Multislice image simulations for one through six atomic layers shown in **d**, and HAADF-STEM images viewed in [103] projection in **e** for several layers and in **f** for near single layer. Scalebars in **e** and **f** are 1 nm.

RESULTS/DISCUSSION

Single crystals of MnPS$_3$ were synthesized through the chemical vapor transport method[40] with a detailed description in the Methods section. Sheets of MnPS$_3$ were directly exfoliated to TEM grids using thermal release tape, omitting a transfer step in order to avoid chemical residues and simplify the process. The crystal structure of MnPS$_3$ is shown in a few different orientations in **Figure 1**. Due to the relative shift of the layers as seen in Figure 1**b**, the [103] orientation is the most natural viewing orientation after cleaving, with the plane of Mn atoms orthogonal to the beam axis. STEM image simulations were performed using the μSTEM package[41] in the same zone axis, and reveal an abundance of atomic contrast variations depending on number of atomic layers, helping to aid in determining specimen thickness. Visualization of the atomic structure is achieved with aberration corrected HAADF-STEM imaging at 60 kV, where panels **e** and **f** in Figure 1 show the stark difference between only a few layers. In this projection, two of the three elements are simultaneously present in each atomic column – either S and Mn or S and P – enabling a variety of contrast at a small number of layers.

The quasi-particle self-consistent *GW* (QS*GW*)[42–44] method combined with the self-consistent solution of Bethe-Salpeter equation (BSE), QS$G\hat{W}$[45–47] is used to calculate the band structure, giving a band gap near 3.6 eV (Supplementary materials), where we note that the more common local-density approximation (LDA) provides a significantly different value (0.5 eV). It is known that QSGW theory overestimates band gap values rather drastically (where LDA can underestimate it), and while QSG$\hat{W}$ greatly reduces this overestimation, it is remains curiously large compared to prior calculations and experiments.

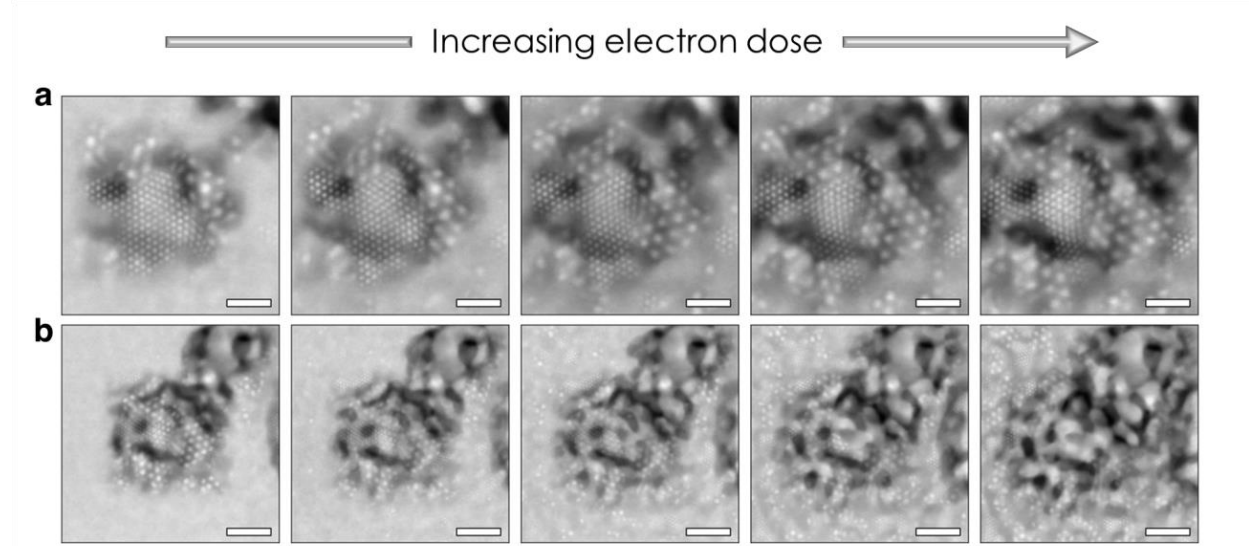

**Figure 2**. Beam-induced mesoscopic ordering. HAADF-STEM images acquired sequentially with a defocused electron probe showing formation of periodic features on the nanometer scale. Two different fields of view (FOV) are shown separately in **a** and **b**. FOV in **a** is 50 nm with 110 seconds of exposure between snapshots, where FOV in **b** is 100 nm with 516 seconds of exposure between snapshots. With a nominal 20 pA beam current throughout the entire experiment, an electron dose of $5.5 \times 10^4 \ e^-/\text{Å}^2$ and $6.4 \times 10^4 \ e^-/\text{Å}^2$ for **a** and **b** respectively. Videos and frames of smaller time steps found in supplementary materials. Scalebars in **a** and **b** are 10 and 20 nm, respectively.

During routine HAADF-STEM imaging, we observe that in several layer MnPS$_3$, the electron beam begins to modify the structure locally. This modification can include degradation of materials and formation of holes for large exposures, as well as much more subtle changes of atomic structure. Experimentally, we have established that optimal conditions for observation of this dynamic is to intentionally defocus the converged electron beam on the order of 50 nm and sweep the beam in a field of view of at least several nanometers. Different scan strategies (paths, dwell times, defocus, etc.) were tested, and while most any condition causes these dynamics, the slightly defocused approach appeared to increase the rate of formation of periodic features, we presume because the dose is balanced with the probe size under this defocused condition. In other words, a slightly defocused probe provides enough dose to locally induce changes but not too much to destroy the structure too quickly.

If a large field of view of the specimen is inspected with a moderate imaging duration using the described defocused approach, well-ordered structures start to emerge – these are shown in several time increments in **Figure 2**, with complete videos and tracking of the phases found in the Supplementary materials. One likely will immediately recognize such periodic lattice modulations that are reminiscent of the moiré interference seen in twisted materials in both STM and STEM.[48,49] We note that several apparent moiré wavelengths appear, and in a way compete with one another leading to a rather spectacular dynamic evolution. This visually striking beam-induced transformation generates ordered periodic structures on the mesoscopic scale, but questions remain as to what causes the periodic behavior and what properties the modulated structures exhibit.

It is important to note that the formation of these periodic patterns is often followed by significant damage in the material and formation of holes. However, the process can be controlled and stopped at different stages, and often very complex dynamics of a periodically ordered phase is observed prior to the onset of irreversible damage. This observation strongly suggests the potential role of beam-induced stoichiometry changes in the emergence of these structures. We find no such beam-induced ordering effects in the sister material $MnPSe_3$ but do observe typical damage (likely Se vacancy formation) that agrees with other work on MoS2 and MoSe2,[50,51] where frame by frame analysis can be found in Supplementary materials. Both $MnPS_3$ and $MnPSe_3$ were cleaved and baked overnight prior to microscopy, hence hydrocarbon contamination is significantly reduced for these experiments. With these considerations, effects of contamination can be ruled out as having an influence on the observations of mesoscopic ordering in $MnPS_3$. Very recently laser irradiation on a set of $WS_2$/$WS_2$ vdW layers was found to cause sulfur vacancy formation leading to twisting of layers due to the resulting lattice mismatch,[22] which strongly supports our observations when using the electron beam, but in a significantly more localized region. Further, we are able to capture crystal dynamics of the moiré phases due to the atom-sized electron probe.

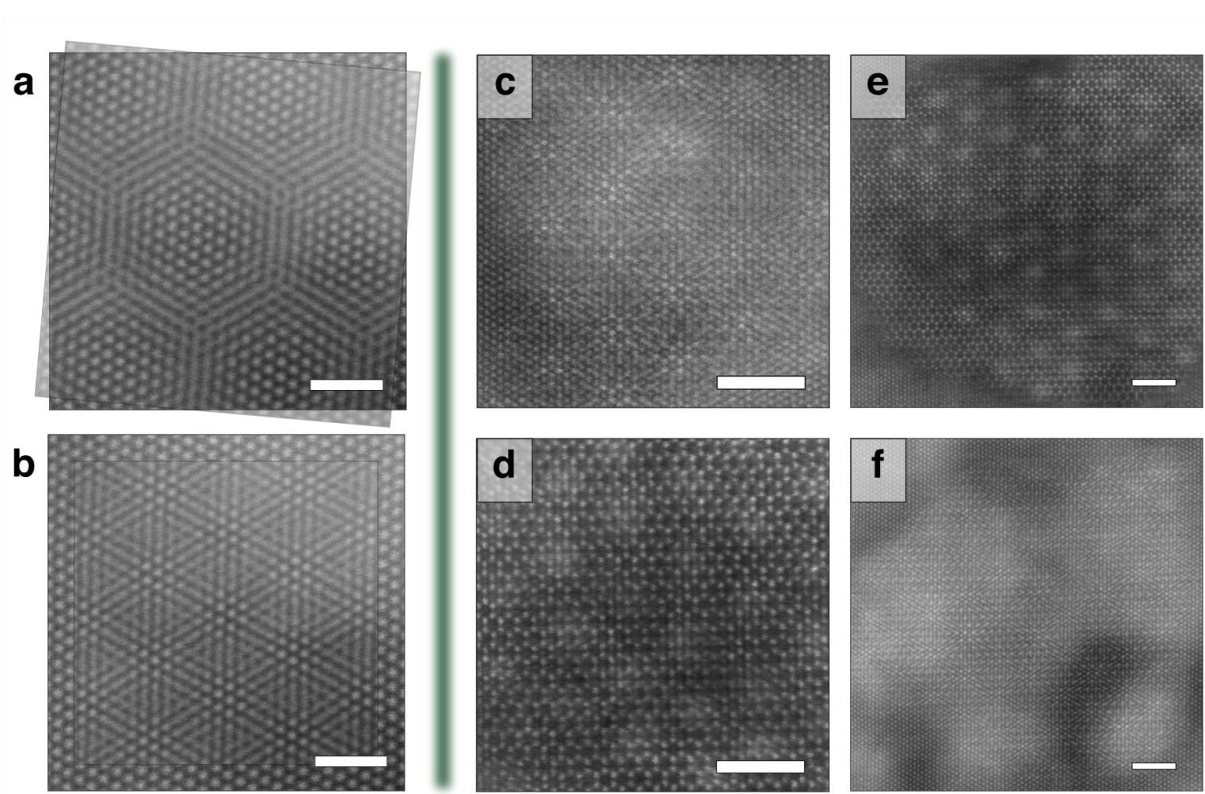

**Figure 3**. Atomistic mechanisms of mesoscopic ordering. Qualitative illustration for emergence of the structures via synthetic 5° rotation (**a**) and 15% compression (**b**) of one image relative to another. Atomic resolution HAADF-STEM imaging shows presence of periodic nanometer scale features at a variety of number of layers in **c**-**f**. All scalebars are 2 nm.

We further explore the qualitative origins of the observed contrast after electron irradiation. After the structures have formed, the electron probe is focused, and small fields of view are captured by HAADF-STEM imaging and shown in **Figure 3**. Arguably the most common circumstance which produces moiré superlattices is when one lattice twists relative to another. If this is synthetically done *via* image analysis techniques as shown schematically in Figure 3**a**, the recognizable moiré features arise in the lattice, but these are distinctly different from what is observed. On the other hand, if one lattice is compressed or expanded relative to the other, a much better qualitative match is seen (Fig 3**b**). This however is a peculiar notion since the lattice compression of one layer that is needed to realize the same degree of long-range period that is observed is on the order of 15% or higher, which is by no means a small amount. Moreover, the transformation is *local*, suggesting the presence a sharp, strained boundary between the native and compressed regions. At the same time, the local nature of transformations implies the ability to pattern such structures and is evaluated later.

While moiré structures form under electron beam irradiation in the presence of any number of layers more than one, a relationship between number of layers and observable configurations is seen in the HAADF-STEM images in Figure 3**c**-**f**. We find that relatively fewer number of layers produce structures shown in Fig 3**e** while greater number of layers produce that shown in Fig 3**f**. Notably, the boundary between the original and newly formed structures is clearly seen in most of the experimental micrographs in Figure 3.

To gain a better understanding of the structure changes, we study the time dynamics of the system under irradiation by collecting stacks of HAADF-STEM images. Both the moiré formation at the mesoscale and the lattice periodicity determined by Fast Fourier Transforms (FFTs) are analyzed at each step. Surprisingly, a large range of periods is observed in the FFT calculated from near the end of the image stack as seen in **Figure 4**, indicating the generation of long-range and short-range periodicity. The indicated outermost lattice planes in Figure 4**b** belong to those of the $3\bar{1}\bar{1}$ family, with a calculated spacing of 1.7516 Å, corresponding well to the smallest periodicity in the system (about 1.71 Å) that exists in the [103] projection throughout the entire experiment. The remaining FFT peaks that arise throughout electron exposure cannot be assigned to a lattice plane in the pristine layered structure. Emergence of the moiré interference occurs after only a few frames of exposure. Multiple moiré phases simultaneously exist – one with ~1.35 nm period and another with ~1.6 nm period; this is not surprising after seeing a variety of order competition in Figure 2. FFT analysis also shows the early formation of a 0.3 nm periodicity which later separates into periods of 0.27 nm and 0.34 nm near 20 frames of exposure, suggesting that - in a local environment - one or more layers are stretched in one direction. We defer the discussion on the crystallographic assignments for these spacings until after the proposed mechanism has been discussed.

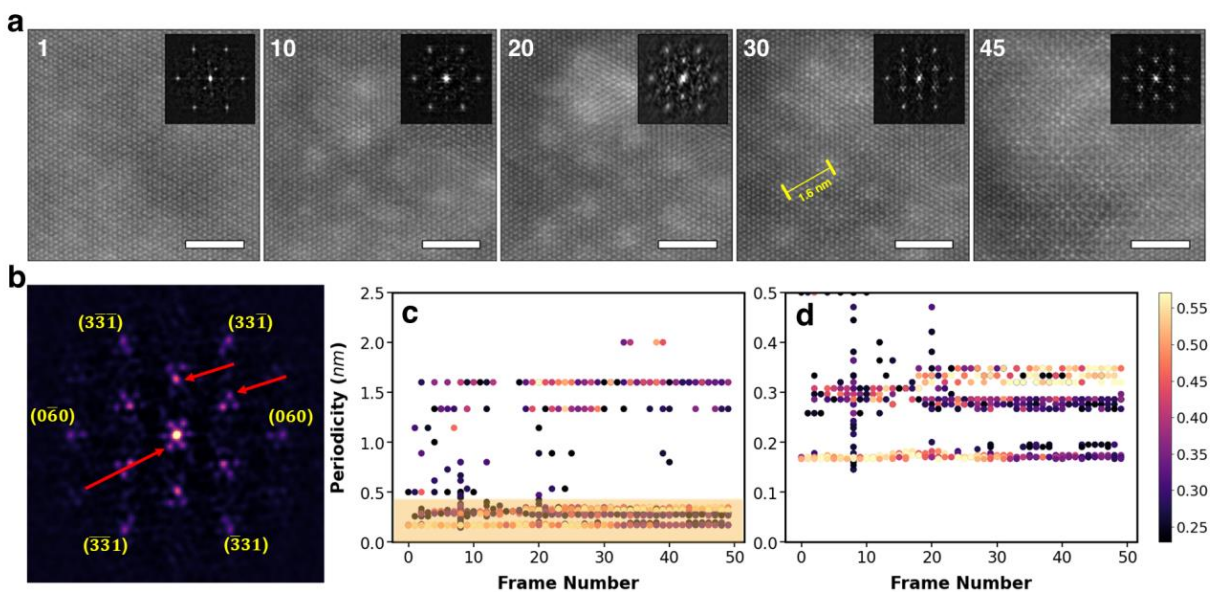

**Figure 4**. Atomistic time dynamics. HAADF-STEM images at specified time intervals (frame numbers) shown in **a**, with FFT insets, where final frame FFT is enlarged in **b** showing emergence of several superlattice peaks and moiré phase (red arrows), in addition to the specified $3\bar{1}\bar{1}$ family

of planes. FFT peaks are detected in each frame and their corresponding real space dimensions are tracked in time in **c**, where the shaded region is blown up in **d**. Scale bars 2 nm.

Our hypothesized mechanism for the ordered transformation is that the electron beam effectively compresses one layer relative to another by removal of sulfur atoms in one layer, producing periodic moiré-like interference – but only where the beam is positioned in the *xy* plane. In other words, the density of vacancies is different within the various layers. Thus, this enables us to generate a type of structure which fundamentally embraces a previously unobserved effect - the properties of which are not yet well understood. We posit that the S atoms are more likely to be removed by knock-on displacement from either the top or bottom layer, or both, such that the internal structure remains intact, but the surfaces are strongly modified. We also suggest that with further electron exposure, internal layers gradually incur vacancies starting from the outermost layers, which may account for the observed mesoscopic order competition of a variety of moiré phases.

In order to understand the origin of observed moiré-like interference, we further explore the role of sulfur vacancies in the lattice using density functional theory (DFT) calculations. **Figure 5** shows that in doing so, the Mn atoms do indeed shift relative to their original atomic coordinates, even with a small number of sulfur atoms removed.

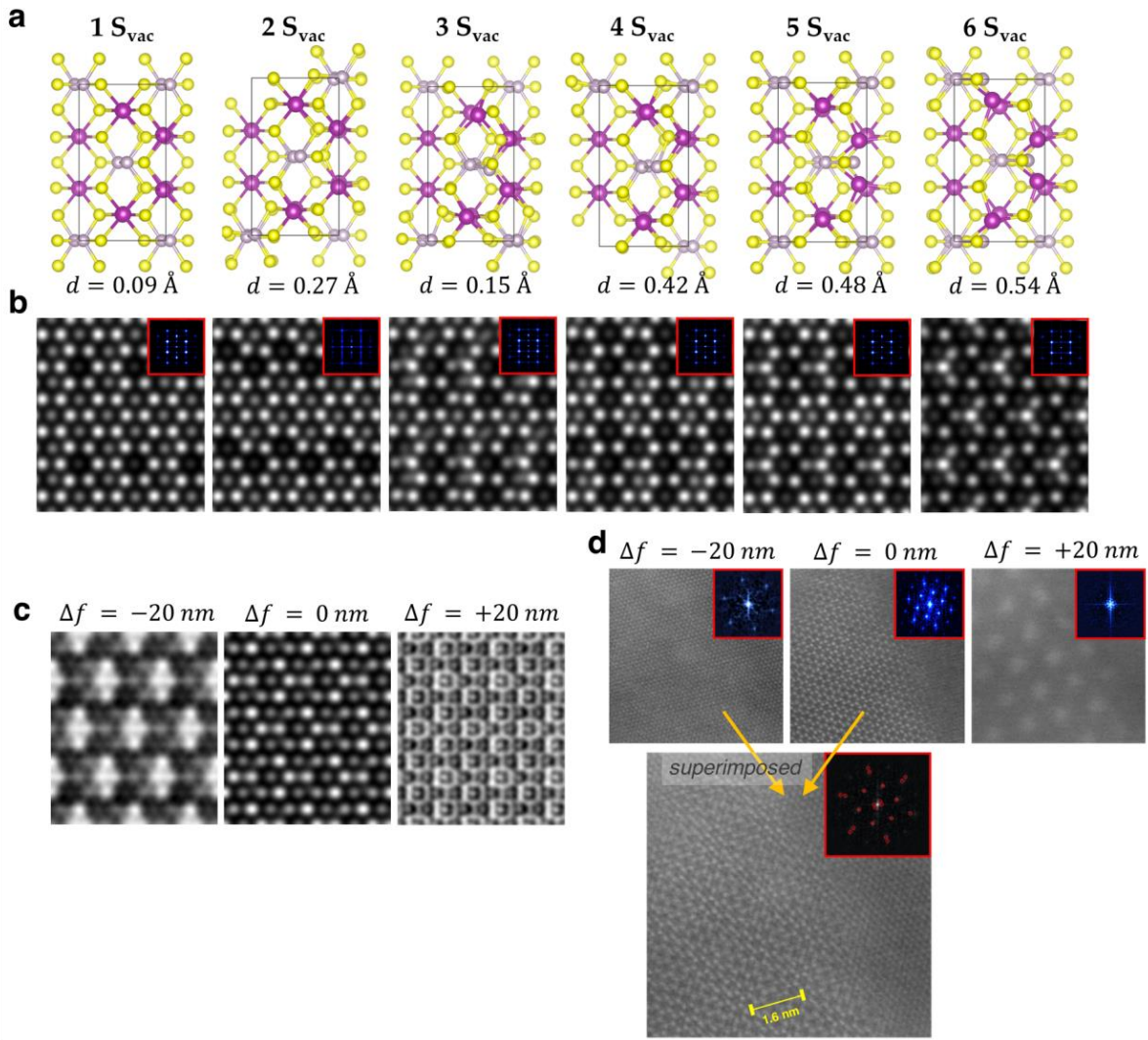

**Figure 5**. Calculated atomic rearrangements upon sulfur vacancy formation. DFT calculations of two-layer system (**a**) where top layer incurs different number of sulfur vacancies, ~~corresponding two-layer system. Where~~ with (**b**) showing the HAADF-STEM image simulations of the corresponding vacancy layers. Note that apparent missing atoms in (**a**) is merely an artifact of the atomic relaxation such that some corner P atoms (for 2 $S_{vac}$ and 4 $S_{vac}$) have shifted outside the unit cell and are therefore no longer visible. Image simulations for arrangement of one layer containing six vacancies positioned at the end of a stack of six pristine layers (relative to the e-beam propagation direction) is shown in (**c**) for varying levels of defocus. Panel (**d**) shows experimental images acquired at selected defocus values, superimposition of two of them and its FFT for comparison. Magnitude of largest atomic displacement, *d*, is provided for each variant in (**a**).

In the DFT approach, we considered a MnPS$_3$ system of two constituent layers. Systematically, sulfur atoms were removed from the top layer. Several configurations of the S vacancies in the vicinity of the Mn were considered. As the lowest-energy sites occur in the neighborhood of Mn, all possible first and second neighboring S vacancy sites were considered. The side views of the representative model indicating the Mn displacement and S vacancy sites are shown in Figure 5**a**.

It is evident from Figure 5 that S vacancies result in a displacement in the stacking of the $MnPS_3$ homo-bilayer. This high-symmetry stacking (AA type) corresponds to regions where one Mn atom is eclipsed by another Mn of the opposite layer in each unit cell. Furthermore, with increasing S vacancy concentration, the stacking of Mn atoms in different layers undergoes a relative increasing displacement which results in moiré-like interference. It is also observed that the displacement of the Mn atoms increases from 0.09 Å (for one S vacancy) to 0.54 Å (full, six-vacancy incorporation). The displacements shown in the six-vacancy schematic agree with our FFT analysis in that the hexagon of Mn atoms (with six S vacancies) undergoes an anisotropic "stretching" that results in apparent superlattice peaks.

To verify these assertions, a set of STEM image simulations using the μSTEM package were performed, first of individual vacancy layers (Figure 5**b**), then of a stack of layers consisting of six pristine layers followed by a single six-vacancy layer (Figure 5**c**), for select values of defocus. A more in-depth study of image simulations at more defocus values can be found in the Supplemental materials. These results indicate there is both a strong dependence on the number of sulfur vacancies as well as defocus value, and we imagine even more variability if multiple layers are simultaneously considered with different vacancy density. The defocused image simulation for -20 nm shows the emergence of longer-range periodicity of about 0.5 nm. This does - albeit briefly - match the time series periodicity shown in Figure 4 (near frame 3). Supplementary materials show additional time series recordings where the 0.5 nm periodicity persists for a slightly longer duration, but in all experimental cases, this period vanishes, and almost immediately longer-range periods emerge. Based on these considerations, we conclude that the 0.5 nm periodicity observed in the defocused image simulations is effectively an intermediate phase toward longer-range moiré periods that approach upwards of 1.5 and 2 nm which form as a result of additional vacancy formation in multiple layers. We note that the moiré periods seen in Figure 4 are observed in multiple experiments where the Supplementary materials show these in detail. Not surprisingly, a strong dependence on the defocus value is also found experimentally in a focal series stack, and moiré interference is clearly observed by superimposing two different image sections in the stack shown in Figure 5**d**. The period observed by the FFT of the superimposed sections shows a 1.6 nm moiré period, but no peaks represent 0.5 nm as in the image simulation, suggesting at this stage we already observe the interference between a set of pristine layers and multiple vacancy layers – not just one.

Returning to Figure 4, one can now understand that the 0.3 nm periodicity emerges due to one layer incurring several sulfur vacancies. We attribute this 0.3 nm FFT peak to the 200 set of lattice planes in a structure where the bottom layer of the crystal contains six sulfur vacancies (as shown in Figure 5**a**) and the remaining six layers are pristine $MnPS_3$. Since many possible moiré phases can exist due to a variety of combinations of pristine and vacancy layers, it is difficult to exactly match image simulations with each phase. Despite this, the emergence of the shorter-range 0.3 nm period clearly results when one vacancy layer is formed at the end of a pristine system, and this remains in the system for the remainder of the experiment, though it does separate into 0.34 nm and 0.27 nm periods presumably meaning at least two different vacancy layers are now present – and this is roughly the same time the longer-range periods appear.

We next consider whether this modification can modify the electronic and dielectric properties of the material, as was recently demonstrated in $MoS_2$/$WSe_2$ twisted heterostructures,[52] providing a route for feature engineering. In the STEM environment, EELS provides a direct link to the electromagnetic local density of states[53,54] on the nanometer and even atomic scales. The incident electron energy, here $E_0 = 60 kV$, has a probability to transfer a portion of its energy inelastically that leads to an electronic excitation event, where the change in electron energy is detected using a magnetic spectrometer. STEM-EELS allows to probe the local electronic behavior by collecting the EEL spectra at an array of probe positions – this is performed here for both a region of moiré structures and a larger region containing step edges of several layers. By nature, the data structure is three dimensional (two of space, one of energy), hence visualization can be non-intuitive. For this reason, non-negative matrix factorization (NMF)[55] – a spectral decomposition technique that is now common in the electron microscopy and other fields[56–58] – is chosen to explore the data space and is presented in **Figure 6**, in addition to showing true EEL spectra from a few selected locations.

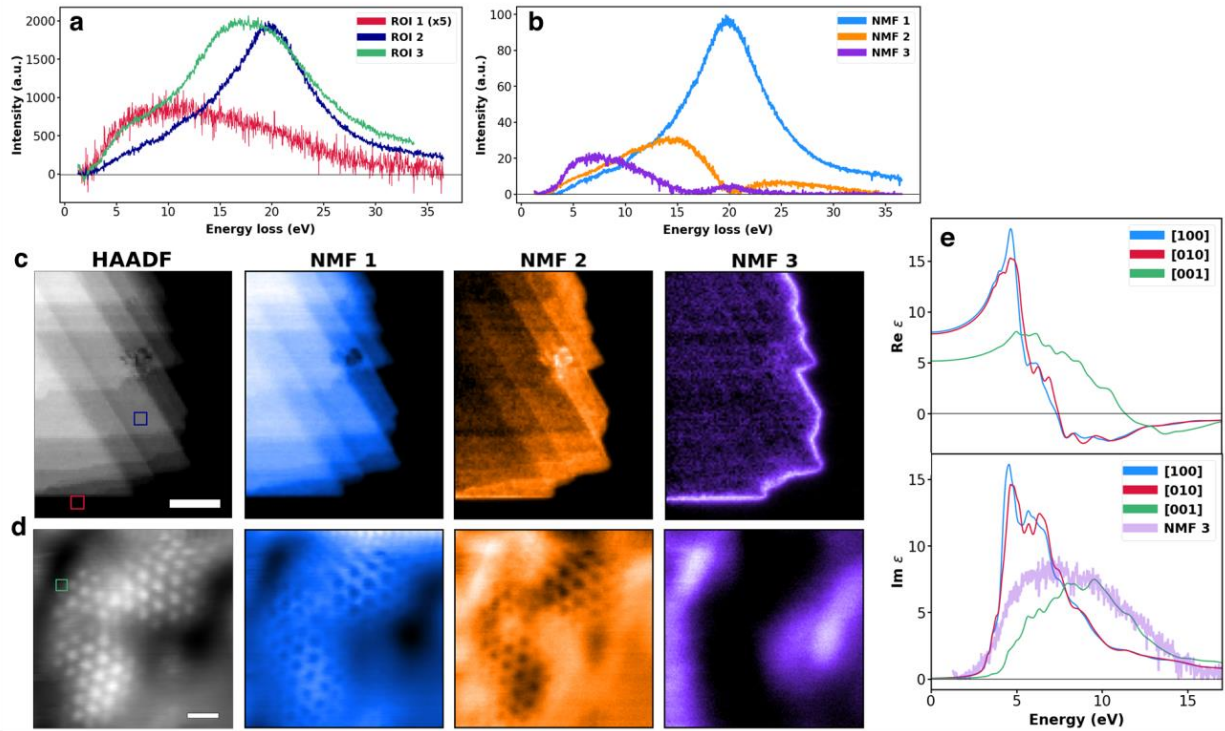

**Figure 6**. STEM-EELS mapping and computed dielectric response. EEL spectra from selected regions within corresponding HAADF-STEM images are shown in **a**; NMF decomposition is used to explore the primary spectral features where spectral endmembers shown in **b**. Large scale native features in panel **c**; electronic behavior arising from moiré features in **d**. Data in **c** are decomposed

by three-component NMF extraction; NMF model from **c** is used to decompose hyperspectral data in **d**. Real and imaginary parts of the in-plane and out-of-plane dielectric response are shown in **e**, with NMF 3 component overlaid for comparison. Scalebars in **c** and **d** are 50 nm and 5 nm, respectively.

Examination of a 2D flake on the mesoscopic scale shows several distinct plasmonic features seen in Figure 6**a-d**. We recently discovered the presence of a highly localized edge plasmon in MnPS$_3$ *via* autonomous active learning;[59] here, localization and response of this plasmon are clearly seen in the NMF 3 component (purple) whose peak energy lies near 6 eV. The length scale of the edge plasmon is curiously on the same order as the emergent mesoscopic ordering, suggesting the possibility of coupling between the two effects. Next, we inspect the energy loss variations arising within the beam-induced moiré regions, again by using NMF to reveal if there exists any underlying plasmonic discrepancies from formation of the structures. Edge plasmons are generated as evidenced by NMF 3 component in Figure 6**d**, which is not surprising given that the electron beam has created a new edge by removal of some material. The bulk resonance reduces in energy with decreasing number of layers and can be used as a fingerprint in thickness determination.[26] In this way, NMF generally separates thin from thick regions which is clearly seen in Figure 6**c** (NMF 1 and NMF 2). The averaged EEL spectra from selected regions of interest (ROI) shown in Figure 6**a** show features that are similar to those revealed by NMF. The discrepancy in peak energy position between ROI2 and ROI3 does not necessarily imply this is due to the moiré formation, but instead is likely due to a difference in thickness between these two specimens which can cause a shift in the bulk plasmon energy. A subtle change manifests in the plasmon resonance near 15 eV (Fig. 6**d**, NMF 2) where the relative intensity decreases in the moiré regions and increases immediately outside them, in contrast to the resonance near 20 eV which behaves as expected where the intensity reduces for thinner regions and is constant otherwise. This suggests that there is indeed a change to the local plasmon response upon mesoscopic ordering and with further study could be a basis for coupling to edge modes. Additionally, there is a small region in Figure 6**c** where the beam has induced local ordering and NMF 2 component is most intense in this region. The macroscopic dielectric response is computed incorporating the electron-hole two-particle correlations through higher order Feynman diagrams. The calculated dielectric properties of MnPS$_3$ offer insight into potential utility in the plasmonic arena. The real and imaginary parts of the dielectric function presented in Figure 6**e** clearly show that the out-of-plane absorption is dramatically different from in-plane absorption. A more detailed comparison of calculations can be found in the supplemental materials. We note that in EELS, the electron beam is considered to be an unpolarized source, therefore the spectrum represents a combination of both in-plane and out-of-plane components of the permittivity. Particularly, the total absorption can be correlated with certain electronic features contained within the energy loss spectrum, both of which clearly exhibit a broad response from about 4 to 12 eV, apparent in Figure 6**e** when comparing NMF component 3. While the electron source in an electron microscope cannot utilize the strong anisotropy in the dielectric behavior, polarized optical sources certainly can with ease, and therefore MnPS$_3$ has an even stronger potential in future nanophotonic applications.

Finally, the feasibility of locally patterning the mesoscopically ordered structures is evaluated in **Figure 7**. Edge engineering[6] in transition metal dichalcogenides using the electron beam in the STEM has been ongoing for several years, hence we envision patterning of $MnPS_3$ and possibly other metal phosphorous trichalcogenides to complement these efforts. Initial testing of simple rectangular shapes directly utilizing rectangular sub-scans yields the structures seen in Figure 7**a**. Taking this a step further we realized more sophisticated patterns can be generated, such as the laboratory logo seen in Figure 7**b**. Even here, several different moiré phases have emerged. Automating the formation of these structures is therefore critical, and while we show the first steps towards this goal, implementation of feedback into the patterning routines undoubtedly will lead to more precise structures, increased reproducibility, and targeted functionalities.

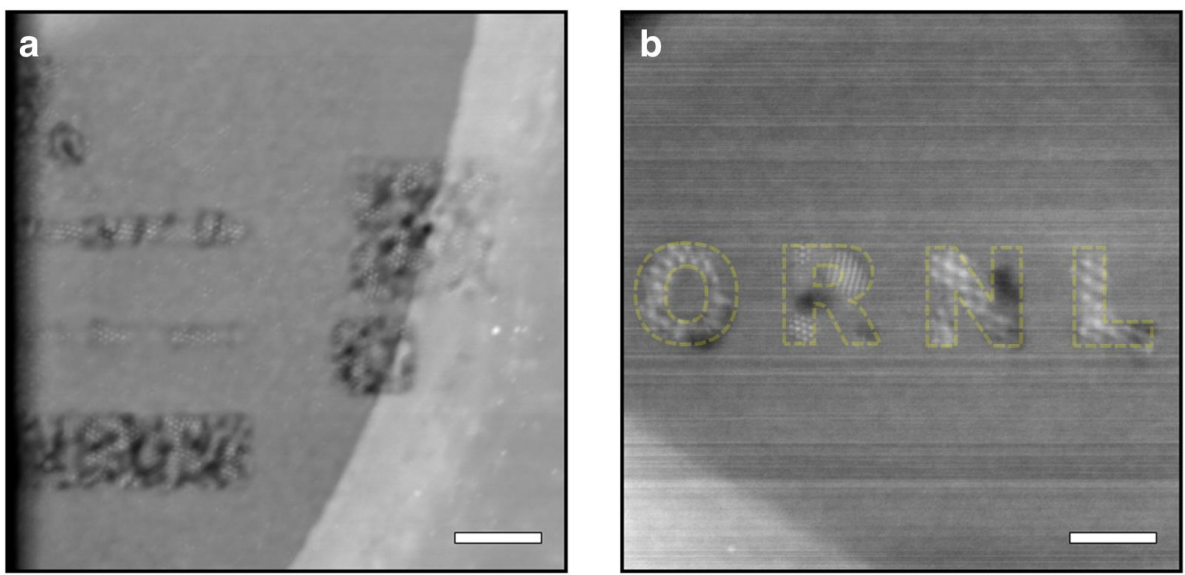

**Figure 7**. Patterning of mesoscopic ordered phenomena. Simple rectangular features in **a**; more complex shapes in **b**. Scale bars in **a** and **b** are 50 and 20 nm.

CONCLUSIONS

To summarize, we report highly unusual structural reconstructions in layered $MnPS_3$ associated with the emergence of the distinct mesoscale periodicities. Unlike the twisted structures, in this case the mechanism seems to be associated with the expansion/contraction of individual layers, the behavior we attribute to beam-induced stoichiometry changes. Critically, these behaviors are induced and controlled by the electron beam, demonstrating a broad range of dynamic behaviors resembling the nucleation and growth of different phases, and allowing for selective control of matter on the atomic scales. We further explore the evolution of the plasmonic properties induced by the formation of these ordered structures.

These findings provide a fundamentally distinct route towards engineering the electronic and quantum properties locally, using the $MnPS_3$ and similar materials as a beam-controlled component. The fact that the many stacking patterns in the metal phosphorous trichalcogenide materials are nearly iso-energetic leads us to speculate that such effects may also occur in other metal phosphorous trichalcogenide compounds (those with sulfur) allowing for an even greater number of beam-engineerable materials waiting to be discovered. Recent progress in autonomous microscopy[59–61] and other automated experiments[62] may provide exactly this pathway. These materials coupled with beam engineering are promising candidates for beyond CMOS devices.

**Acknowledgements:** This effort (electron microscopy) is based upon work supported by the U.S. Department of Energy (DOE), Office of Science, Basic Energy Sciences (BES), Materials Sciences and Engineering Division (K.M.R., S.V.K.) and was performed and partially supported at the Oak Ridge National Laboratory's Center for Nanophase Materials Sciences (CNMS), a U.S. Department of Energy, Office of Science User Facility. This research used resources of the Compute and Data Environment for Science (CADES) at the Oak Ridge National Laboratory, which is supported by the Office of Science of the U.S. Department of Energy under Contract No. DE-AC05-00OR22725. V.S. acknowledges the XSEDE allocation (Grant No. TG- DMR200008) and the Infrastructure for Scientific Applications and Advanced Computing (ISAAC) at the University of Tennessee for the computational resources. Research at the University of Tennessee is supported by the U.S. DOE, BES, Materials Science Division under award DE-FG02-01ER45885 (J.L.M.). N.H. and D.M. were supported by the National Science Foundation through grant number DMR-1808964. MIK and SA are supported by the ERC Synergy Grant, project 854843 FASTCORR (Ultrafast dynamics of correlated electrons in solids). DP is supported by the National Renewable Energy Laboratories. SA, DP and MIK acknowledge PRACE for awarding us access to Juwels Booster and Cluster, Germany.

METHODS/EXPERIMENTAL

**Growth and exfoliation**

MnPS$_3$ single crystals were synthesized through the chemical vapor transport (CVT) method. Manganese powder (Alfa Aesar 99.95%), phosphorus powder (Alfa Aesar 99.995%) and sulfur chunks (Puratronic 99.9995%) of a ratio 1: 1: 3.1 were thoroughly mixed and ground under argon atmosphere. The mixture was then pressed into a pellet and sealed in a fused tube under vacuum. Polycrystalline powder was obtained after annealing the ampoule in a muffle furnace for one week at 730°C. 2 grams of annealed powder was transferred into a new fused tube. Dehydrated iodine was added as a transport agent. The tube was then connected to a vacuum station and sealed at a proper length. Large single crystal could be harvested at the cold end of the tube after annealing at 700°C for 6 days in a tube furnace.

Samples for electron microscopy were prepared using standard mechanical exfoliation techniques where flakes were directly exfoliated onto Au Quantifoil grids using thermal release tape, skipping transfer steps to and from silicon which eliminates any chemical processes. Only sheets suspended over perforated holes in the membrane were inspected.

**Electron microscopy**

Microscopy was performed using a NION monochromated fifth order aberration corrected scanning transmission electron microscope (MACSTEM). Prior to loading into the microscope, samples were baked overnight in vacuum at 160º C to minimize contamination. An accelerating voltage of 60 kV with nominal probe current of 20 pA was used in the experiments. Pressure at the sample was kept at 10$^{-9}$ Torr or better, minimizing etching effects. Convergence semi-angle was set to a nominal 30 mrad.

**Image simulations**

All simulations were performed using the quantum excitation of phonons option in the $\mu$STEM package.[41] The unit cells were tiled to for a supercell of approximately 3nm by 3 nm and 1024 pixels were used in each direction. An aberration free probe with an aperture of 30 mrad with an energy of 60keV was used. A total of 80 phase gratings were calculated, resulting in a total of 1200 distinct phase grates due to phase ramp shifting of the potentials. Eighty passes through the sample were used in all cases. An annular ADF detector with a 61 mrad inner angle and a 300 mrad outer angle were used. To account for temporal incoherence a focal series of 21 images were calculated covering a range of 10 nm with side of zero defocus. Assuming a chromatic aberration coefficient of $C_c = 1.3 mm$ and a Gaussian energy spread with a fullwidth half maximum (FWHM) of 0.35 eV, the weighted average image was calculated. To account for source size effects, the images were blurred with a Gaussian with a FWHM of 0.075 nm.

**Density functional theory**

The DFT calculations were performed using the Vienna *ab initio* simulation package (VASP) code, using the Perdew–Burke–Ernzerhof (PBE)[63] generalized gradient approximation (GGA), the projector-augmented wave approach, and a cutoff energy of 500 eV for the plane wave expansion of the wave functions. Geometries were relaxed using a conjugate gradient algorithm until the forces on all unconstrained atoms became smaller than < 0.03 eV/Å. Appropriate Monkhorst-Pack k-point meshes were considered to produce results with an energy convergence of 0.5 meV/atom.

**QS$G\hat{W}$ theory**

The quasi-particle self-consistent GW (QS$GW$) method[42–44] is an approach to find an optimal starting point independent, ab-initio independent-particle Hamiltonian self-consistently within the GW approximation to the self-energy ($\Sigma$). The QS$G\hat{W}$[45,46,64] method additionally includes vertex corrections in the polarizability (precursor to the effective interaction W) by solving the Bethe-Salpeter equation (BSE) for the two-particle Hamiltonian in each self-consistency cycle.

The crucial difference in our implementation of QS$G\hat{W}$ from most other implementations of BSE is that our calculations are self-consistent in both self energy $\Sigma$ and the charge density.[64,65] W is calculated with ladder-BSE corrections and the self energy, using a static vertex in the BSE. G, $\Sigma$ and W are updated iteratively until all of them converge. Brillouin zone (BZ) sampling for single particle quantities (density and energy bands with the static quasiparticlized QS$G\hat{W}$ self-energy $\Sigma^0(k)$) used a 6×6×4 origin centered mesh while the (relatively smooth) dynamical self-energy $\Sigma(k)$ was constructed using a 3×3×2 k-mesh using the $\Gamma$ offset points method and $\Sigma^0(k)$ extracted from it. For each iteration in the QS$G\hat{W}$ self-consistency cycle, the charge density was made self-consistent. The QS$G\hat{W}$ cycle was iterated until the RMS change in $\Sigma^0$ reached $10^{-5}$ Ry. Thus the calculation was self-consistent in both $\Sigma^0(k)$ and the density. Numerous checks were made to verify that the self-consistent $\Sigma^0(k)$ was independent of starting point, for both QS$GW$ and QS$G\hat{W}$ calculations; e.g. using LDA or Hartee-Fock self-energy as the initial self energy for QS$GW$ and using LDA or QS$GW$ as the initial self-energy for QS$G\hat{W}$. LDA underestimates the electronic correlations severely in MnPS$_3$ and predicts a band gap of 0.5 eV, QS$GW$ enhances the gap to 4 eV and ladders in QS$G\hat{W}$ screens the W to reduce the gap to 3.6 eV. We check convergence in the QS$G\hat{W}$ band gap and macroscopic dielectric response by increasing the size of the two-particle Hamiltonian. We find that they converge with 60 valence bands and 48 conduction bands. Excitons with positive binding energies ~0.7 eV are observed in the vertex corrected optical spectrum. For our simulations we used the crystal structure id-8613 in materialsproject.org.

**Data availability**

Electron microscopy data and analysis codes are freely available here: https://github.com/kevinroccapriore/MnPS3

The content of the Supporting Information Available online includes individual HAADF-STEM images as well as image stacks (movies) that capture the dynamics of the electron beam induced transformation of MnPS$_3$ at both atomic scale and mesoscopic scales. Also included are STEM image simulations, 3D EELS mapping, experimental and simulated STEM focal series, comparison with MnPSe$_3$, and electronic structure and dielectric property calculations. Analysis of all data is additionally part of this repository.

All the input file structures and the command lines to launch calculations are rigorously explained in the tutorials available on the Questaal webpage.[66] The source codes for LDA, QS$GW$ and QS$G\hat{W}$ are available from Ref [[66]] under the terms of the AGPLv3 license.